\documentclass[conference]{IEEEtran}

\usepackage{graphicx}  

\newcommand{\eeq}{\end{equation}}

\newcommand{\br}{\mbox{\boldmath $r$}}

\newcommand{\bs}{\mbox{\boldmath $s$}}

\newcommand{\bb}{\mbox{\boldmath $b$}}

\newcommand{\bM}{\mbox{\boldmath $M$}}

\newcommand{\bP}{\mbox{\boldmath $P$}}

\newcommand{\bH}{\mbox{\boldmath $H$}}

\newcommand{\bS}{\mbox{\boldmath $S$}}

\newcommand{\bU}{\mbox{\boldmath $U$}}

\newcommand{\bLambda}{{\bf \Lambda}}

\newcommand{\bu}{\mbox{\boldmath $u$}}

\newcommand{\bn}{\mbox{\boldmath $n$}}

\newcommand{\bd}{\mbox{\boldmath $d$}}

\newcommand{\bI}{\mbox{\boldmath $I$}}

\newcommand{\bD}{\mbox{\boldmath $D$}}

\newcommand{\ds}{\displaystyle}

\newcommand{\K}{{\cal K}}
\newcommand{\G}{{\cal G}}
\newcommand{\N}{{\cal N}_0}
\def\R{{\cal R}}
\def\S{{\cal S}}

\newcommand{\beq}{\begin{equation}}

\begin{document}

\title{Non-cooperative Games for Spreading Code Optimization, Power Control and Receiver \\ Design in
Wireless Data Networks}

\author{\authorblockN{Stefano Buzzi}
\authorblockA{Universit\`a degli Studi di Cassino \\
DAEIMI - Via G. Di Biasio, 43\\
I-03043 Cassino (FR), Italy\\
Email: buzzi@unicas.it}
\and
\authorblockN{H. Vincent Poor}
\authorblockA{School of Engineering and Applied Science\\
Princeton University\\
Princeton, NJ, 08544, USA \\
Email: poor@princeton.edu}}

%


\maketitle

\begin{abstract}
This paper focuses on the issue of energy efficiency in wireless data networks through a game theoretic approach. The case considered
is that in which each user is allowed to vary its transmit power, spreading code, and uplink receiver in order to maximize its own utility,
which is here defined as the ratio of data throughput to transmit power. In particular,  the case in which linear multiuser detectors are employed at the receiver
is treated first, and, then, the more challenging case in which non-linear decision feedback multiuser receivers are adopted is addressed.
It is shown that, for both receivers, the problem at hand of utility maximization can be regarded as a non-cooperative game, and it is proved
 that a unique Nash equilibrium point exists. Simulation results show that significant performance gains can be obtained through
both non-linear processing and spreading code optimization; in particular, for systems with a number of users not larger than
 the processing gain, remarkable gains come from spreading code optimization, while, for overloaded systems, the largest gains
come from the use of non-linear processing. In every case, however, the non-cooperative games proposed  here are shown to
outperform competing alternatives.
\end{abstract}


%
\IEEEpeerreviewmaketitle

\section{Introduction}
Game theory \cite{gtbook} is a branch of mathematics that has been applied primarily in economics and other social sciences to study the interactions among several autonomous subjects with contrasting interests. More recently, it has been discovered that it can also be used for the design and analysis of communication systems, mostly with application to resource allocation algorithms \cite{gt}, and, in particular, to power control \cite{yates}.
As examples, the reader is referred to \cite{nara1,nara2,SaraydarPhD}. Here, for a multiple access wireless data network,
noncooperative and cooperative games are introduced, wherein each user chooses its  transmit power in order to maximize its own utility, defined as the ratio of the throughput to transmit power.
While the above papers consider the issue of power control assuming that a conventional matched filter is available at the receiver, the recent paper \cite{meshkati} considers the problem of joint linear receiver design and power control so as to maximize the utility of each user. It is shown here that the inclusion of receiver design in the considered game brings remarkable advantages, and, also, results based on the powerful large-system analysis are presented.

This paper is the first in this area that considers the cross-layer issue of utility maximization with respect to the choice
of receiver, spreading code and transmit power. First of all, we generalize the game considered in \cite{meshkati} by considering also spreading code optimization. We show that iterative algorithms, of the same kind proposed in \cite{ulukusyener}, can be applied to our scenario in order to improve the achieved Signal-to-Noise plus Interference (SINR) of each user. We will show that the newly considered noncooperative game admits a unique Nash equilibrium and achieves remarkable gains with respect to the performance levels attained by the solution proposed in \cite{meshkati}.
Then, we consider the problem of utility maximization with respect to transmit power and spreading code, for the case in which a {\em non-linear} decision feedback receiver is used. We thus propose two noncooperative games wherein first transmit power is chosen so as to maximize utility, and, then, joint spreading code optimization and power control is undertaken for utility maximization.
Our results will show that remarkable gains are granted by the use of spreading code optimization when the number of users does not exceed the processing gain (undersaturated region), while, for saturated systems, non-linear interference cancellation, eventually coupled with code optimization, provides the most significant gains.

The rest of this paper is organized as follows. The next section contains some preliminaries and the system model of interest. Section III introduces a non-cooperative game for the case in which linear receivers are employed, while in Section IV we introduce two non-cooperative utility maximization games for the case that a non-linear interference cancellation receiver is adopted. In Section V we present and discuss the results of some computer simulations that show the merits of the proposed games and their advantages with respect to competing alternatives. Finally, we give some concluding remarks in Section VI.

\section{Preliminaries and Problem Statement}
Consider the uplink of a $K$-user synchronous, single-cell, direct-sequence code division multiple access (DS/CDMA) network with processing gain $N$ and subject to flat fading. After chip-matched filtering and sampling at the chip-rate, the $N$-dimensional received data vector, say $\br$, corresponding to one symbol interval, can be written as
\beq
\br=\ds \sum_{k=1}^{K}\sqrt{p_k} h_k b_k \bs_k + \bn \; ,
\label{eq:r}
\eeq
wherein $p_k$ is the transmit power of the $k$-th user\footnote{To simplify subsequent notation, we assume that the transmitted power $p_k$ subsumes also the gain of the transmit and receive antennas.}, $b_k\in \{-1,1\}$ is the information symbol of the $k$-th user, and $h_k$ is the real\footnote{We assume here, for simplicity, a real channel model; generalization to practical channels, with I and Q components, is straightforward.} channel gain between the $k$-th user's transmitter and the access point (AP); the actual value of $h_k$ depends on both the distance of the $k$-th user's terminal from the AP and the channel fading fluctuations. The $N$-dimensional vector $\bs_k$ is the spreading code of the $k$-th user; we assume that the entries of $\bs_k$ are real and that $\bs_k^T \bs_k=\|\bs_k\|^2=1$, with $(\cdot)^T$ denoting transpose. Finally, $\bn$ is the ambient noise vector, which we assume to be a zero-mean white Gaussian random process with covariance matrix $(\N/2) \bI_N$, with $\bI_N$ the identity matrix of order $N$. An alternative and compact representation of (\ref{eq:r}) is given by
\beq
\br=\bS \bP^{1/2}\bH \bb + \bn \; ,
\label{eq:r2}
\eeq
wherein $\bS=[ \bs_1, \ldots, \bs_K]$ is the $N\times K$-dimensional spreading code matrix, $\bP$ and $\bH$ are $K \times K$-dimensional diagonal matrices, whose diagonals are $[p_1, \ldots, p_K]$ and $[h_1, \ldots, h_K]$, respectively, and, finally, $\bb=[b_1, \ldots, b_K]^T$ is the $K$-dimensional vector of the data symbols.

Assume now that each mobile terminal sends its data in packets of $M$ bits, and that it is interested both in having its data received with as small as possible error probability at the AP, and in making  careful use of the energy stored in its battery. Obviously, these are conflicting goals, since error-free reception may be achieved by increasing the received SNR, i.e. by increasing the transmit power, which of course comes at the expense of battery life\footnote{Of course there are many other strategies to lower the data error probability, such as for example the use of error correcting codes, diversity exploitation, and implementation of optimal reception techniques at the receiver. Here, however, we are mainly interested to energy efficient data transmission and power usage, so we consider only the effects of varying the transmit power, the receiver and the spreading code on energy efficiency.}. A useful approach to quantify these conflicting goals is to define the utility of the $k$-th user as the ratio of its throughput, defined as the number of information bits that are received with no error in unit time, to its transmit power \cite{nara1,nara2}, i.e.
\beq
u_k=\ds \frac{T_k}{p_k}\; .
\label{eq:utility}
\eeq
Note that $u_k$ is measured in bit/Joule, i.e. it represents the number of successful bit transmissions that can be made for each Joule of energy drained from the battery.
Denoting by $R$ the common rate of the network (extension to the case in which each user transmits with its own rate $R_k$ is quite simple) and assuming that each packet of $M$ symbols contains $L$ information symbols and $M-L$ overhead symbols, reserved, e.g., for channel estimation and/or parity checks, the throughput $T_k$ can be expressed as
\beq
T_k=\ds R \frac{L}{M} P_k
\label{eq:Tk}
\eeq
wherein $P_k$ denotes the the probability that a packet from the $k$-th user is received error-free. In the considered DS/CDMA setting, the term $P_k$ depends formally on a number of parameters such as the spreading codes of all the users and the diagonal entries of the matrices $\bP$ and $\bH$, as well as on the strength of the used error correcting codes. However, a customary approach is to model the multiple access interference as a Gaussian random process, and assume that $P_k$ is an increasing function of the $k$-th user's Signal-to-Interference plus Noise-Ratio (SINR) $\gamma_k$, which is naturally the case in many practical situations.

Recall that, for the case in which a linear receiver is used to detect the data symbol $b_k$, according, i.e., to the decision rule
\beq
\widehat{b}_k=\mbox{sign}\left[\bd_k^T \br\right] \; ,
\label{eq:decrule}
\eeq
with $\widehat{b}_k$ the estimate of $b_k$ and $\bd_k$ the $N$-dimensional vector representing the receive filter for the user $k$, it is easily seen that the SINR $\gamma_k$ can be written as
\beq
\gamma_k=\ds \frac{p_k h_k2 (\bd_k^T \bs_k)2}{\frac{\N}{2}\|\bd_k\|^2 + \ds \sum_{i \neq k} p_i h_i2
(\bd_k^T \bs_i)2} \; .
\label{eq:gamma}
\eeq
Of related interest is also the mean square error (MSE) for the user $k$, which, for a linear receiver, is defined as
\beq
{\rm MSE}_k= E \left\{ \left(b_k - \bd_k^T \br \right)2 \right\}=1 + \bd_k^T \bM \bd_k - 2\sqrt{p_k} h_k \bd_k^T
\bs_k \; ,
\eeq
wherein $E\left\{ \cdot \right\}$ denotes statistical expectation and $\bM=\left(\bS \bH \bP \bH^T \bS^T + \frac{\N}{2} \bI_N\right)$ is the covariance matrix of the data.

\medskip

The exact shape of $P_k(\gamma_k)$ depends  on factors such as the modulation and coding type.
However, in all cases of relevant interest, it is an increasing function of $\gamma_k$ with a sigmoidal shape, and converges to unity as $\gamma_k \rightarrow + \infty$; as an example, for  binary phase-shift-keying (BPSK) modulation coupled with no channel coding, it is easily shown that
\beq
P_k(\gamma_k)=\left[1-Q(\sqrt{2\gamma_k})\right]^M \; ,
\label{eq:psr}
\eeq
with $Q(\cdot)$ the complementary cumulative distribution function of a zero-mean random Gaussian variate with unit variance. A plot of  (\ref{eq:psr}) is shown in Fig. 1 for the case $M=100$.

It should be noted that substituting (\ref{eq:psr}) into (\ref{eq:Tk}), and, in turn, into (\ref{eq:utility}), leads to a strong incongruence. Indeed, for $p_k \rightarrow 0$, we have $\gamma_k \rightarrow 0$, {\em but} $P_k$ converges to a small but non-zero value (i.e. $2^{-M}$), thus implying that an unboundedly  large utility can be achieved by transmitting with zero power, i.e. not transmitting at all and making blind guesses at the receiver on what data were transmitted. To circumvent this problem, a customary approach \cite{nara2,meshkati} is to replace $P_k$ with an {\em efficiency function}, say $f_k(\gamma_k)$, whose behavior should approximate as close as possible that of $P_k$, except that for $\gamma_k \rightarrow 0$ it is required that $f_k(\gamma_k)= o(\gamma_k)$. The function
$f(\gamma_k)=(1-e^{-\gamma_k})^M$ is a widely accepted substitute for the true probability of correct packet reception, and in the following we will adopt this model\footnote{See Fig. 1 for a comparison between the
Probability $P_k$ and the efficiency function.}.
This efficiency function is increasing and S-shaped, converges to unity as $\gamma_k$ approaches infinity, and has a continuous first order derivative. Note that we have omitted the subscript $``k''$, i.e. we have used the notation $f(\gamma_k)$ in place of $f_k(\gamma_k)$ since we assume that the efficiency function is the same for all the users.

Summing up, substituting (\ref{eq:Tk}) into (\ref{eq:utility}) and replacing the probability $P_k$ with the above defined efficiency function, we obtain the following expression for the $k$-th user's utility:
\beq
u_k=R \ds \frac{L}{M} \frac{f(\gamma_k)}{p_k} \; , \quad \forall k=1, \ldots, K \; .
\label{eq:utility2}
\eeq

Now, based on the utility definition (\ref{eq:utility2}), many interesting questions arise concerning how each user may maximize its utility, and how this maximization affects utilities achieved by other users. Likewise, it is natural to question what happens in a non-cooperative setting wherein each user autonomously and selfishly tries to maximize its own utility, with no care for other users utilities. In particular, in this latter situation, is the system able to reach an equilibrium wherein no user is interested in varying its parameters since each action it would take would lead to a decrease in its own utility?  Game theory  provides means to study these interactions and to provide some useful and insightful answers to these questions.

Initially, game theory was applied in this context mainly as a tool to study non-cooperative scenarios wherein mobile users are allowed to vary their transmit power only (see \cite{nara1,nara2,SaraydarPhD}, for example) to maximize utility, and where conventional matched filtering is used at the receiver. Recently, instead, in  \cite{meshkati} such an approach has been extended to the cross layer scenario in which each user may vary its power and its uplink {\em linear} receiver, i.e. the problem of joint linear multiuser detection optimization and power control for utility maximization has been tackled. In the following, we will go further by considering the case of spreading code choice, power control and {\em linear} receiver design for utility maximization. Moreover, the case in which a parametric non-linear decision feedback receiver is used will be considered, and new games wherein optimization of this receiver, spreading code choice and power control is performed jointly in order to maximize utility will be proposed.

\begin{figure}
\centering
\includegraphics[width=7cm]{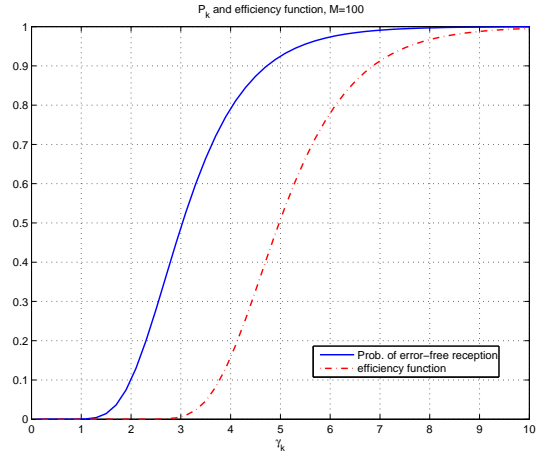}
\caption{Comparison of probability of error-free packet reception and efficiency function versus receive SINR and for packet size $M=100$. Note the S-shape of both functions.}
\label{figampiezza}
\end{figure}

\section{Non-cooperative games with linear receivers}
We begin by considering a noncooperative game wherein each user aims to maximizing its own utility by varying its spreading code, its transmit power, and its linear uplink receiver.
Formally, the proposed game $\G$ can be described as the triplet $\G=\left[\K, \left\{\S_k\right\}, \left\{u_k\right\} \right]$, wherein $\K= \left\{1, 2, \ldots, K\right\}$ is the set of active users participating in the game,
$u_k$ is the $k$-th user's utility defined in (\ref{eq:utility2}), and
\beq
\S_k=[0, P_{k,\max}] \times \R^N \times \R_1^N \; ,
\label{eq:strategy}
\eeq
is the set of possible actions (strategies) that user $k$ can take. It is seen that $\S_k$ is written as the Cartesian product of three different sets, and indeed $[0, P_{k, \max}]$ is the range of available transmit powers for the $k$-th user (note that $P_{k, \max}$ is the maximum allowed transmit power for user $k$), $\R^N$, with $\R$ the real line, defines the set of all possible linear receive filters, and, finally,
$$
\R_1^N= \left\{\bd \in \R^N \; : \; \bd^T \bd=1 \right\} \; ,
$$
defines the set of the allowed spreading codes\footnote{Here we assume that the spreading codes have real entries; the problem of utility maximization with reasonable complexity for the case of discrete-valued entries is a challenging issue that will be considered in the future.} for user $k$.

Before proceeding further, it is also convenient to define the concept of {\em Nash equilibrium}. Let
$$
(s_1, s_2, \ldots, s_K) \in \S_1 \times \S_2 \times \ldots \S_K
$$
denote a certain strategy $K$-tuple for the active users. The point $(s_1, s_2, \ldots, s_K)$ is a Nash equilibrium if for any user $k$, we have
$$u_k(s_1, \ldots, s_k, \ldots, s_K) \ge u_k(s_1, \ldots, s_k^*, \ldots, s_K)\; ,
$$
$\forall s_k^* \neq s_k\, .$ Otherwise stated, at a Nash equilibrium, no user can {\em unilaterally} improve its own utility by taking a different strategy. A fast reading of this definition might lead to think that at Nash equilibrium users' utilities achieve their maximum values. Actually, this is not the case, since the existence of a Nash equilibrium point does not imply that no other strategy $K$-tuple does exist that can lead to an improvement of the utilities of some users while not decreasing the utilities of the remaining ones. These latter strategies are usually said to be Pareto-optimal \cite{gtbook}.
Otherwise stated, at a Nash equilibrium, each user, provided that the other users' strategies do not change, is not interested in changing its own strategy. However, if some sort of cooperation would be available, users might agree to simultaneously switch to a different strategy $K$-tuple, so as to improve the utility of some, if not all, active users. In this paper, we will focus on Nash equilibrium points only, since they are the result of non-cooperative games. Moreover, it can be shown, although this is not discussed here due to lack of space, that, for the considered problem, the utilities achieved by Nash-equilibrium points are only slightly smaller than those achieved on the Pareto-optimal frontier of the game.

Summing up, the proposed noncooperative game can be cast as the following maximization problem
\beq
\ds \max_{\S_k} u_k = \max_{p_k, \bd_k, \bs_k} u_k(p_k, \bd_k, \bs_k ) \; , \quad
\forall k=1, \ldots, K \; .
\label{eq:game}
\eeq
Given  (\ref{eq:utility2}), the above maximization can be also written as
\beq
\ds \max_{p_k, \bd_k, \bs_k} \frac{f(\gamma_k(p_k, \bd_k, \bs_k))}{p_k} \; , \quad
\forall k=1, \ldots, K \; .
\eeq
Moreover, since the efficiency function is monotone and non-decreasing, we also have
\beq
\ds \max_{p_k, \bd_k, \bs_k} \frac{f(\gamma_k(p_k, \bd_k, \bs_k))}{p_k}= \max_{p_k}  \frac{f\left(\ds \max_{\bd_k, \bs_k}\gamma_k(p_k, \bd_k, \bs_k)\right)}{p_k} \; ,
\label{eq:deriv}
\eeq
i.e. we can first take care of SINR maximization with respect to spreading codes and linear receivers, and then focus on maximization of the resulting utility with respect to transmit power.

With regard to this latter point, recall that, if a linear Minimum MSE (MMSE) receiver is used, the following relation can be shown to hold \cite{wangpoor}
\beq
{\rm TMSE}= \ds \sum_{i=1}^K {\rm MSE}_i = \ds \sum_{i=1}^K \frac{1}{1+ \gamma_i} \; ,
\label{eq:sirmse}
\eeq
wherein TMSE is the total MSE. Otherwise stated, among linear receivers, the MMSE receiver is the one that maximizes the SINR vector $(\gamma_1, \ldots, \gamma_K)$. Now assume that we wish to minimize the MSE for each user by varying not only the receiver, but also the spreading code. This problem has been considered in \cite{ulukusyener,ensuring,rose}; in particular, if we let $\bD=[\bd_1, \ldots, \bd_K]$ and denote by $(\cdot)^+$ Moore-Penrose pseudoinversion, it has been therein shown that the sum of the MSE's of all the users admits a unique global optimum, and that the iterations
\beq
\begin{array}{lll}
\bd_i=\sqrt{p_i} h_i \left(\bS \bH \bP \bH^T \bS^T + \frac{\N}{2} \bI_N\right)^{-1}\!\! \bs_i   & \forall i=1, \ldots, K \\
\bs_i=\sqrt{p_i} h_i \left(p_i h_i2 \bD \bD^T + \mu_i \bI_N \right)^{+} \bd_i  & \forall i=1, \ldots, K
\end{array}
\label{eq:iterazioni}
\eeq
admit a unique stable fixed point that is the global minimizer of the total MSE. In the above relations, $\mu_i$ should be set so that $\|\bs_i\|=1$. No details are given in \cite{ulukusyener} on how this could be done in an efficient way, so in the Appendix we outline a procedure  for  efficiently finding the value of $\mu_i$ ensuring the constraint $\| \bs_i \|=1$.

Now, it is natural to ask if minimization of the total MSE with respect to both linear receivers and spreading codes still maximizes the user's SINR's. We can thus state the following result.

\noindent
 {\bf Lemma 1:}
{\em Let $\bar{\bS}$ and $\bar{\bD}$ be the spreading code matrix and the linear receiver matrix that jointly achieve the global minimum of the total MSE. Then, no strategy of spreading
codes and decoder can be found to increase the SINR of one or more users without decreasing the SINR of at least one other user.
}
\\
{\bf Proof:} If $\bar{\bS}$ and $\bar{\bD}$ are the global minimizers of the MSE, then $\bar{\bD}$ contains the MMSE receivers resulting from the spreading codes of $\bar{\bS}$. Denote by $\left\{ \gamma_i(\bar{\bS}, \bar{\bD})\right\}_{i=1}^K$ the SINR values achieved by the matrices $\bar{\bS}$ and $\bar{\bD}$. Assume now that there exists a spreading code matrix $\bS^* \neq \bar{\bS}$ such that $\gamma_i({\bS}^*, \bar{\bD}) > \gamma_i(\bar{\bS}, \bar{\bD})$, for at least one $i \in \{1, \ldots, K\}$ and
$\gamma_j({\bS}^*, \bar{\bD}) \geq \gamma_j(\bar{\bS}, \bar{\bD})$ for $j \neq i$. If this is the case, we can make an MMSE update and obtain the matrix $\bD^*$ of the MMSE receivers corresponding to the codes in $\bS^*$. For a given set of spreading codes, using the MMSE receiver always yields a maximization of the SINR and a minimization of the MSE. We thus have
$
\gamma_i(\bS^* , \bD^*)
 > \gamma_i(\bar{\bS}, \bar{\bD})$, and
$
\gamma_j(\bS^* , \bD^*)
 \geq \gamma_j(\bar{\bS}, \bar{\bD}) $, $\forall j \neq i$.
Consequently, given relation (\ref{eq:sirmse}), we have
$$
{\rm TMSE}(\bS^* , \bD^*) < {\rm TMSE} (\bar{\bS}, \bar{\bD}) \; ,
$$
which contradicts the starting assumptions that $\bar{\bS}$ and $\bar{\bD}$ are the global minimizers of the MSE.
\hfill \rule{2mm}{2mm}

\medskip

We are now ready to express our result on the non-cooperative game for spreading code optimization, linear receiver design and power control.

\noindent
 {\bf Proposition 1:}
{\em The non-cooperative game defined in (\ref{eq:game}) admits a unique Nash equilibrium point $(p_k^*, \bd_k^*, \bs_k^*)$,
for $k=1, \ldots, K$, wherein
\begin{itemize}
\item[-]
$\bs^*_k$ and $\bd^*_k$ are the unique $k$-th user
spreading code and receive filter\footnote{Actually the linear receive filter is unique up to a positive scaling factor.} resulting from iterations (\ref{eq:iterazioni}). Denote by $\gamma_k^*$ the corresponding SINR.
\item[-]
$p_k^*=\min \{\bar{p}_k, P_{k, \max} \}$, with $\bar{p}_k$ the $k$-th user transmit power such that the $k$-th user maximum SINR $\gamma_k^*$ equals $\bar{\gamma}$, i.e. the unique solution of the equation $f(\gamma)=\gamma f'(\gamma)$, with $f'(\gamma)$ the derivative of $f(\gamma)$.
\end{itemize}}
\noindent
{\bf Proof:} The proof is a generalization of the one provided in \cite{meshkati} and so is only briefly sketched here.
Since $\partial \gamma_k /\partial p_k=\gamma_k /p_k$, it is easily seen that each user's utility is maximized if each user is able to achieve the SINR $\bar{\gamma}$, that is the unique\footnote{Uniqueness of $\bar{\gamma}$ is
ensured by the fact that the efficiency function is S-shaped \cite{rodriguez}.} solution of the equation $f(\gamma)=\gamma f'(\gamma)$. By Lemma 1, running iterations (\ref{eq:iterazioni}) until convergence is reached provides the set of spreading codes and MMSE receivers that maximize the SINRs for all the users. As a consequence, the utility of each user is maximized by adjusting transmit powers so that the optimized (with respect to spreading codes and linear receivers) SINRs equal $\bar{\gamma}$.

So far, we have shown how to  set the transmit power, spreading code and receiver design to maximize utility at the Nash equilibrium. It remains to be shown that a Nash equilibrium exists. Luckily, we can use the same arguments of \cite{nara2} and state that a unique Nash equilibrium point exists since each user's utility function is quasi-concave\footnote{A function is quasi-concave if there exists a point below which the function is
nondecreasing, and above which the function is nonincreasing.} in the transmit power $p_k$ and since the efficiency function is S-shaped.
\hfill \rule{2mm}{2mm}

\section{Non-cooperative games with nonlinear decision-feedback receivers}

Consider now the case in which a non-linear decision feedback receiver is used at the receiver.  We assume that the users are indexed according to a non-increasing sorting of their channel gains, i.e. we assume that $h_1 > h_2 > \ldots, h_K$. We consider a serial interference cancellation (SIC) receiver wherein detection of the symbol from the $k$-th user is made according to the following rule
\beq
\widehat{b}_k=\mbox{sign}\left[\bd_k^T \left(\br - \sum_{j<k} \sqrt{p_j} h_j \widehat{b}_j \bs_j\right) \right\} \; .
\label{eq:decruleSIC}
\eeq
Otherwise stated, when detecting a certain symbol, the contribution from the data symbols that have been already detected is subtracted from the received data. If past decisions are correct, users that are detected later enjoy a considerable reduction of multiple access interference, and indeed the SINR for user $k$, under the assumption of correcteness of past decisions, is written as
\beq
\gamma_k=\ds \frac{p_k h_k2 (\bd_k^T \bs_k)2}{\frac{\N}{2}\|\bd_k\|^2 + \ds \sum_{j > k} p_j h_j2
(\bd_k^T \bs_j)2} \; .
\label{eq:gammaSIC}
\eeq
A considerable amount of literature exists on  decision feedback receivers, and many detectors of this kind have been proposed and analyzed. Here, our goal is just to show that non-linear receivers coupled with spreading code optimization and power control can bring remarkable performance advantages with respect to linear receivers. As a consequence, we consider only the decision rule (\ref{eq:decruleSIC}) and introduce noncooperative games built on that, with no further optimization. As an example, receiver (\ref{eq:decruleSIC}) might be optimized with respect to the users' detection order, or by using properly distorted versions of the signal to be subtracted; these issues will not be considered here due to lack of space.

Now, given receiver (\ref{eq:decruleSIC}) and the SINR expression (\ref{eq:gammaSIC}), we consider the problems of utility maximization with respect to the transmit power, spreading code choice, and receivers $\bd_1, \ldots, \bd_K$. To begin with, let us neglect spreading code optimization and consider the problem
\beq
\ds \max_{p_k, \bd_k} \frac{f(\gamma_k(p_k, \bd_k))}{p_k} \; , \quad
\forall k=1, \ldots, K \; .
\label{eq:gameSIC}
\eeq

The following result can be shown to hold.

\noindent
 {\bf Proposition 2:}
{\em Define $\bS_k=[\bs_{k}, \ldots, \bs_K]$, $\bP_k=\mbox{diag}(p_{k}, \ldots, p_K)$ and
$\bH_k=\mbox{diag}(h_{k}, \ldots, h_K)$.
The non-cooperative game defined in (\ref{eq:gameSIC}) admits a unique Nash equilibrium point $(p_k^*, \bd_k^*)$,
for $k=1, \ldots, K$, wherein
\begin{itemize}
\item[-]
$\bd^*_k=\sqrt{p_k} h_k(\bS_k \bH_k \bP_k \bH_k^T \bS_k^T + \frac{\N}{2}\bI_N)^{-1} \bs_k$
is the unique $k$-th user
receive filter\footnote{Uniqueness here means up to a positive scaling factor.} that maximize the user $k$ SINR $\gamma_k$ given in (\ref{eq:gammaSIC}). Denote $\gamma_k^*=\max_{\bd_k}\gamma_k$.
\item[-]
$p_k^*=\min \{\bar{p}_k, P_{k, \max} \}$, with $\bar{p}_k$ the $k$-th user transmit power such that the $k$-th user maximum SINR $\gamma_k^*$ equals $\bar{\gamma}$, i.e. the unique solution of the equation $f(\gamma)=\gamma f'(\gamma)$, with $f'(\gamma)$ the derivative of $f(\gamma)$.
\end{itemize}}

\noindent
{\bf Proof:} The proof is omitted here due to lack of space. It can be constructed along the same lines as that of Proposition 1. \hfill \rule{2mm}{2mm}

\medskip

Consider, finally, the maximization
\beq
\ds \max_{p_k, \bd_k, \bs_k} \frac{f(\gamma_k(p_k, \bd_k, \bs_k))}{p_k} \; , \quad
\forall k=1, \ldots, K \; .
\label{eq:gameSIC2}
\eeq

The existence and uniqueness of a Nash equilibrium for this game is guaranteed by the following result.

\noindent
{\bf Proposition 3:}
{\em Define $\bD_k=[\bd_{1}, \ldots, \bd_k]$.
The non-cooperative game defined in (\ref{eq:gameSIC2}) admits a unique Nash equilibrium point $(p_k^*, \bs_k^* , \bd_k^*)$,
for $k=1, \ldots, K$, wherein
\begin{itemize}
\item[-]
$\bd^*_k$ and $\bs_k^*$
are the unique stable fixed points of the iterations
$\bd_k=\sqrt{p_k} h_k \left(\bS_k \bH_k \bP_k \bH_k^T \bS_k^T + \frac{\N}{2} \bI_N\right)^{-1}\!\! \bs_k$ and
$\bs_k=\sqrt{p_k} h_k \left(p_k h_k2 \bD_k \bD_k^T + \mu_k \bI_N \right)^{+} \bd_k$,  $\forall k=1, \ldots, K$
and with $\mu_k$ such that $\|\bs_k\|=1$.
Denote by $\gamma_k^*$ the $k$-th user's SINR resulting from the choices $\bs_k=\bs_k^*$ and $\bd_k=\bd_k^*$.
\item[-]
$p_k^*=\min \{\bar{p}_k, P_{k, \max} \}$, with $\bar{p}_k$ the $k$-th user transmit power such that the $k$-th user maximum SINR $\gamma_k^*$ equals $\bar{\gamma}$, i.e. the unique solution of the equation $f(\gamma)=\gamma f'(\gamma)$, with $f'(\gamma)$ the derivative of $f(\gamma)$.
\end{itemize}}

\noindent
{\bf Proof:} The proof is omitted here due to lack of space.  \hfill \rule{2mm}{2mm}

\begin{figure}[t]
\centering
\includegraphics[width=7cm]{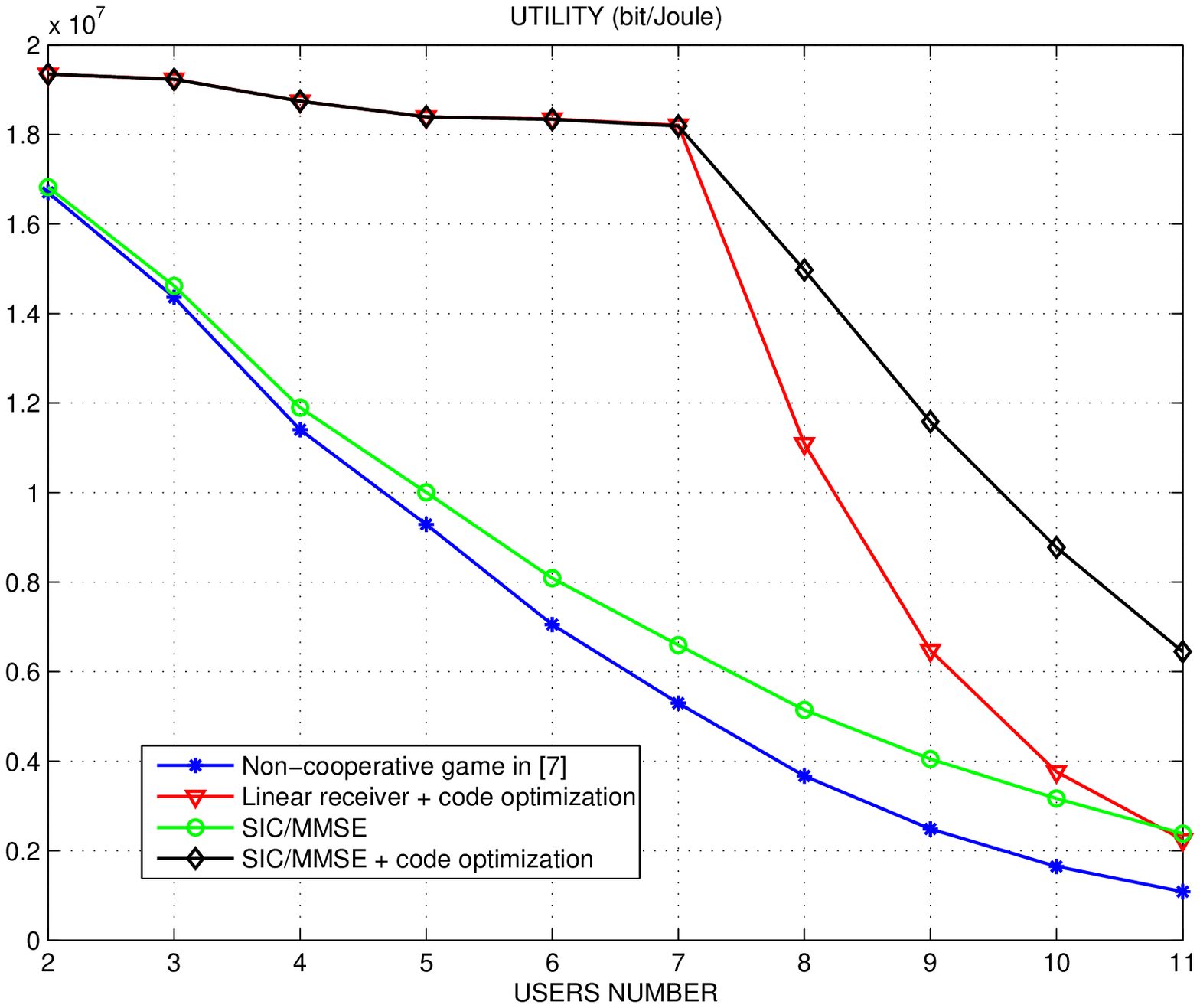}
\caption{Achieved average utility versus number of active users for the proposed noncooperative games and for the game in
reference \cite{meshkati}. The system processing gain is $N=7$.}
\end{figure}

\begin{figure}[t]
\centering
\includegraphics[width=7cm]{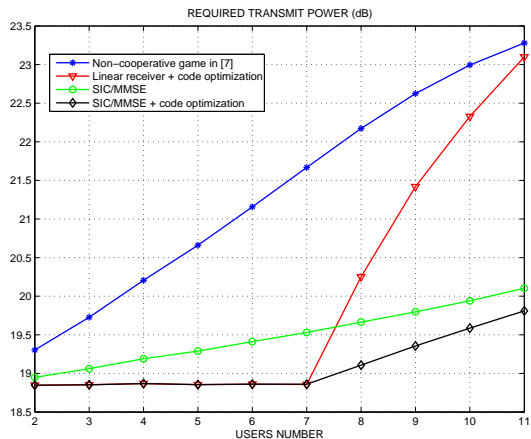}
\caption{Average transmit power versus number of active users for the proposed noncooperative games and for the game in reference \cite{meshkati}. The system processing gain is $N=7$.}
\end{figure}

\section{Numerical Results}
In this section we illustrate some simulation results that give  insight into the performance of the proposed non-cooperative games. We contrast here the performance of the noncooperative game discussed in \cite{meshkati} with that of the games proposed here.
We consider an uplink DS/CDMA system with processing gain $N=7$, and assume that the packet length is $M=120$.
for this value of $M$ the equation $f(\gamma)=\gamma f'(\gamma)$ can be shown to admit the solution $\bar{\gamma}=6.689 = 8.25$dB.
A single-cell system is considered, wherein users may have random positions with a distance from the AP ranging from 10m to 500m. The channel coefficient $h_k$ for the generic $k$-th user is assumed to be Rayleigh distributed with mean equal to $d_k^{-2}$, with $d_k$ being the distance of user $k$ from the AP\footnote{Note that we are here assuming that the power path losses are proportional to the fourth power of the path length, which is reasonable in urban cellular environments.}. We take the ambient noise level to be $\N=10^{-9}$W/Hz, while the maximum allowed power $P_{k,\max}$ is $25$dB. We present the results of averaging over $104$ independent realizations for the users locations, fading channel coefficients and starting set of spreading codes. More precisely, for each iteration we randomly generate an $N \times K$-dimensional spreading code matrix with entries in the set $\left\{-1/\sqrt{N}, 1/\sqrt{N}\right\}$; this matrix is then used as the starting point for the games that include spreading code optimization, and as the spreading code matrix for the games that do not perform spreading code optimization.

Figures  2 - 4 report the achieved average utility (measured in bits/Joule), the average user transmit power and the average achieved SINR at the receiver output for the game in \cite{meshkati} and for the three non-cooperative games considered in this paper. Inspecting the curves, the following conclusions can be drawn. First of all, it is seen that all of the proposed games outperform the one proposed in \cite{meshkati}: of course this result can be easily justified by noting that the proposed games can take advantage of the optimization of the spreading codes and of the superior performance that non-linear receivers provide over linear ones.
As an example, it is seen that for a system with $K=N=7$ the game with SIC/MMSE plus spreading code optimization achieves a utility that is more than 3 times larger than that achieved by the game in \cite{meshkati} and with a simultaneous average transmit power saving of almost 3dB.

For $K\leq N$, a very substantial performance gain can be obtained by resorting to spreading code optimization; indeed, when $K\leq N$, users can be given orthogonal spreading codes, so that the multiaccess channel reduces to a superposition of $K$ separate single-user AWGN channels. Obviously, in this situation the spreading code optimization algorithms converge to a set of orthogonal codes, and this explains the performance gains reported in the figures. Interestingly, for $K\leq N$ the performance of the linear MMSE and of the SIC/MMSE receivers with spreading code optimization coincide: this is an indirect confirmation that in this case the steady-state spreading codes are orthogonal, since in this case no distinction occurs between the SIC/MMSE receiver and the linear one.
In the oversaturated region (i.e. for $K>N$), instead, the merits of the non-linear SIC/MMSE can be clearly seen. Indeed, in this situation the spreading codes, whether optimized or not, are linearly dependent, and this leads to a severe performance degradation for any linear processing. In this region SIC/MMSE plus spreading code optimization is thus the best option, followed by SIC/MMSE with no spreading code optimization. Note that there is a crossing around $K=11$ between the performance of the MMSE receiver with spreading code optimization and that of the SIC/MMSE with no spreading code optimization, revealing that for lightly loaded systems much can be gained through spreading code optimization, while for heavily loaded systems the most significant gains come from the use of non-linear processing. It is also seen from Fig. 4 that receivers achieve on the average an output SINR that is smaller than the target SINR $\bar{\gamma}$: indeed, due to fading and distance path losses, achieving the target SINR would require for some users a transmit power larger than the maximum allowed power $P_{k, \max}$, and so these users are not able to achieve the optimal target SINR. As a confirmation of this, in Fig. 5 we report the fraction of users transmitting at the maximum power: it is seen here that even for the SIC/MMSE receiver with spreading code optimization this fraction is larger than $0.1$.

\begin{figure}[t]
\centering
\includegraphics[width=7cm]{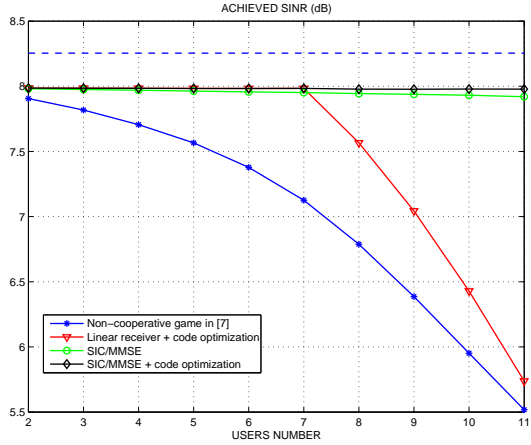}
\caption{Achieved average output SINR versus number of active users for the proposed noncooperative games and for the game in reference \cite{meshkati}. The system processing gain is $N=7$.}
\end{figure}

\begin{figure}[t]
\centering
\includegraphics[width=7cm]{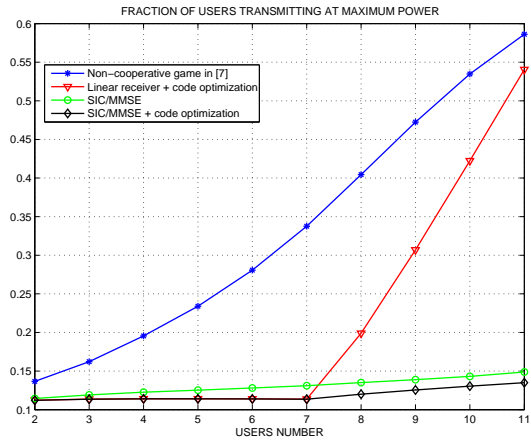}
\caption{Average fraction of users transmitting at their maximum allowed power versus number of active users for the proposed noncooperative games and for the game in reference \cite{meshkati}. The system processing gain is $N=7$.}
\end{figure}

\section{Conclusion}
In this paper the cross-layer issue of joint power control, spreading code optimization and receiver design for wireless data networks has been addressed using a game-theoretic framework. Building on \cite{meshkati},  we have proposed a more general framework wherein also spreading code optimization and non-linear decision feedback multiuser receivers can be used to further increase the energy efficiency of CDMA-based wireless networks.
It has been shown that spreading code optimization in non-overloaded system, and non-linear reception techniques in overloaded systems, bring remarkable performance gains.

The authors' current research is focused on the development and the analysis of adaptive algorithms able to implement the said games without prior knowledge of the fading channel coefficients.

\section*{Acknowledgment}

This research in this paper was supported by the Italian National Research Council (CNR), by the U. S. Air Force Research
Laboratory under Cooperative Agreement No. FA8750-06-1-0252, and by the U. S. Defense Advanced Research Projects
Agency under Grant No. HR0011-06-1-0052. The authors wish to thank Dr. Husheng Li for insightful comments on a preliminary version of this paper.

\section*{Appendix}
Given the relation
\beq
\bs_k=\sqrt{p_k} h_k \left(p_k h_k2 \bD \bD^T + \mu_k \bI_N \right)^{+} \bd_k
\label{eq:app1}
\eeq
we show here how to choose the constant $\mu_k$ so that $\|\bs_k\|=1$. Let $\bU \bLambda \bU^T$ be the eigendecomposition of the matrix $p_k h_k2 \bD \bD^T$. Obviously, $\bU$ is an orthonormal matrix whose  columns are the eigenvectors of $p_k h_k2 \bD \bD^T$, and $\bLambda$ is the corresponding diagonal eigenvalue matrix. Note that some of these eigenvalue will be zero for $K<N$. Now, letting
$\bu_i$ and $\lambda_i$ denote the $i$-th column  of $\bU$ and the $i$-th diagonal element of $\bLambda$, respectively, and
\beq
z(\lambda_i, \mu_k)=\left\{\begin{array}{llll} \frac{1}{\lambda_i + \mu_k} & \quad & {\rm if} \; \lambda_i + \mu_k \neq 0 \\
0 & \quad & {\rm if} \; \lambda_i+\mu_k=0 \; , \end{array} \right.
\eeq
it is easy to show that  (\ref{eq:app1}) can be rewritten as
\beq
\bs_k=\sqrt{p_k} \ds \sum_{i=1}^{N} z(\lambda_i, \mu_k) \bu_i \bu_i^T \bd_k \; .
\label{eq:app2}
\eeq
From (\ref{eq:app2}) it is seen that, as $\mu_k\rightarrow + \infty$, $\|\bs_k\| \rightarrow 0$, thus implying that there exists a finite constant $Q_u$ such that $\|\bs_k\| < 1$ for any $\mu_k\geq Q_u$. Now, let $\lambda_m=\min_i{\lambda_i}$ (note that $\lambda_m$ may be 0 if $K<N$ or in general if $\bD \bD^T$ is not of full rank). It is easy to show that, as $\mu_k \rightarrow \lambda_m^+$, $\|\bs_k\|\rightarrow +\infty$. Accordingly, there exists a finite constant $Q_l >\lambda_m$ such that $\|\bs_k\| > 1$ for $\mu_k \in ]\lambda_m, Q_l]$. Since $\|\bs_k\|$ is monotonically decreasing for $\mu_k \in [Q_l, Q_u]$ and since $\|\bs_k\|>1$ for $\mu_k=Q_l$ and $\|\bs_k\|<1$ for $\mu_k=Q_u$, there exists just one value of $\mu_k$, say $\mu_k^*$, such that $\|\bs_k\|=1$ for $\mu_k=\mu_k^*$. The value of $\mu_k^*$ can be found using standard methods.



\begin{thebibliography}{99}

\bibitem{gtbook}
D. Fudenberg and J. Tirole, {\em Game Theory}, Cambridge, MA: MIT Press, 1991.

\bibitem{gt}
A. B. MacKenzie and S. B. Wicker, ``Game theory in communications: Motivation, explanations, and applications to power control,''
{\em Proc. IEEE Global Telecommun. Conference}, San Antonio, TX, 2001.

\bibitem{yates}
R. D. Yates,
``A framework for uplink power control in cellular radio systems,''
{\em IEEE J. Sel. Areas Comm.}, Vol. 13, pp. 1341-1347, Sep. 1995.

\bibitem{nara1}
D. J. Goodman and N. B. Mandayam, ``Power control for wireless data,'' {\em IEEE Pers. Commun.}, vol. 7, pp. 48-54, Apr. 2000.

\bibitem{nara2}
C. U. Saraydar, N. B. Mandayam and D. J. Goodman, ``Efficient power control via pricing in wireless data networks,''
{\em IEEE Trans. Commun.}, vol. 50, pp. 291-303, Feb. 2002.

\bibitem{SaraydarPhD}
C. U. Saraydar, ``Pricing and power control in wireless data networks,''
Ph.D. dissertation, Dept. Elect. Comput. Eng., Rutgers University, Piscataway, NJ, 2001.

\bibitem{meshkati}
F. Meshkati, H. V. Poor, S. C. Schwartz and N. B. Mandayam,
``An energy-efficient approach to power control and receiver design in wireless data networks,''
{\em IEEE Trans. Comm.}, Vol. 53, pp. 1885-1894, Nov. 2005.

\bibitem{ulukusyener}
S. Ulukus and A. Yener,
``Iterative transmitter and receiver optimization for CDMA networks,''
{\em IEEE Trans. Wireless Commun.}, Vol. 3, pp. 1879-1884, Nov. 2004.

\bibitem{wangpoor}
X. Wang and H. V. Poor,
{\em Wireless Communication Systems: Advanced Techniques for Signal Reception}. Upper Saddle River, NJ: Prentice-Hall, 2004.


\bibitem{ensuring}
P. Anigstein and V. Anantharam, ``Ensuring convergence of the MMSE iteration for interference avoidance to the global optimum,'' {\em IEEE Trans. Inform. Th.}, Vol. 46, pp. 873-885, Sept. 2000.

\bibitem{rose}
C. Rose,
``CDMA codeword optimization: Interference avoidance and convergence via class warfare,'' {\em IEEE Trans. Inf. Th.}, vol. 47, pp. 2368-2382, Sept. 2001.

\bibitem{rodriguez}
V. Rodriguez,
``An analytical foundation for resource management in wireless communication,''
{\em Proc. IEEE Global Telecommun. Conference}, San Francisco, CA, Dec. 2003.


\end{thebibliography}
\end{document}